\documentclass[prc,aps,superscriptaddress,twocolumn,showpacs,nofootinbib]{revtex4}
\usepackage{graphicx,amsmath,amssymb,bm,multirow}
\usepackage{amsthm}
\usepackage{amsfonts,array}
\usepackage{CJK}

\usepackage{dcolumn}
\usepackage{bm}

\usepackage[usenames,dvipsnames]{color}
\usepackage[colorlinks=true,urlcolor=blue,citecolor=blue,breaklinks=true]{hyperref}

\newcommand{\br}{{\mathbf{r}}}

\begin{document}
\begin{CJK}{GBK}{song}

 \preprint{preprint}
\title{Low-energy structure and anti-bubble effect of dynamical correlations in $^{46}$Ar}%
\author{X. Y. Wu}
\affiliation{School of Physical Science and Technology, Southwest University, Chongqing 400715, China}
\author{J. M. Yao}\email{jmyao@swu.edu.cn}
\affiliation{Department of Physics, Tohoku University, Sendai 980-8578, Japan}%
\affiliation{School of Physical Science and Technology, Southwest University, Chongqing 400715, China}
\author{Z. P. Li}
\affiliation{School of Physical Science and Technology, Southwest University, Chongqing 400715, China}

\begin{abstract}
 The low-energy structure of $^{46}$Ar is of particular interest due to the possible coexistence of different shapes and the possible existence of proton ``bubble" structure. In this work, we apply a beyond relativistic mean-field approach to study the low-energy structure of $^{46}$Ar. Correlations beyond the mean field are introduced by configuration mixing of both particle-number and angular-momentum projected axially deformed mean-field states. The low-lying spectroscopy and charge density in laboratory frame are calculated and an excellent agreement with available data is achieved. Even though an evident proton bubble structure is shown in the spherical state of $^{46}$Ar, it eventually disappears after taking into account the dynamical correlation effects. Moreover, our results indicate that the existence of proton bubble structure in argon isotopes is unlikely.
 \end{abstract}

\maketitle



 {\em Introduction}.-The advent of radioactive ion beam facilities and the advances in experimental techniques of gamma-ray spectroscopy have allowed to measure the low-lying spectroscopy of neutron-rich nuclei. Many new phenomena have been revealed in these nuclei. One of the most important findings is the erosion of traditional $N=20$ shell and onset of large collectivity in $^{32}$Mg, which possesses a very low excitation energy of $2^+_1$ state~\cite{Detraz79,Mueller84} and a large $B(E2:0^+_1\to 2^+_1)$  value~\cite{Motobayashi95}. In recent years, the low-energy structure of $^{46}$Ar is of particular interest due to the possible development of deformation and shape coexistence related to the weakening of $N=28$ shell below $^{48}$Ca inferred from the $\beta$-decay experiment~\cite{Sorlin93}, which is however confronted with some controversial indications from other experimental measurements. The neutron single-particle energies determined via the $d$($^{46}$Ar, $^{47}$Ar)$p$ reaction indicate a slightly weakening of $N=28$ shell~\cite{Gaudefroy06}. The systematics of $B(E2: 0^+_1 \to 2^+_1)$ values and $2^+_1$ energies suggest the persistence of $N=28$ shell in $^{46}$Ar~\cite{Scheit96,NNDC}. Recently, the lifetime measurement of $^{46}$Ar by means of the differential recoil distance Doppler shift method results in an increase in $B(E2: 0^+_1 \to 2^+_1)$  from $^{44}$Ar to $^{46}$Ar~\cite{Mengoni10}, which supports the weakening of $N=28$ shell in $^{46}$Ar.

 Meanwhile, the possible existence of proton bubble structure in $sd$-shell nuclei~\cite{Todd-Rutel2004,Grasso2009} has made the ground state of $^{46}$Ar very interesting. With modern self-consistent mean-field approaches restricted to spherical symmetry, the formation of proton bubble resulting from the depopulation of 2s$_{1/2}$ orbital has been predicted in $^{46}$Ar~\cite{Khan08,Chu10} and in some very neutron-rich Ar isotopes~\cite{Khan08}. It was pointed out that pairing correlation effect would quench significantly the bubble structure in $^{46}$Ar, but not for the very neutron-rich $^{68}$Ar~\cite{Khan08}. Since the bubble structure in $^{46}$Ar is sensitive to the occupancy of 2s$_{1/2}$ orbital and therefore to the underlying shell structure, the prediction turns out to be model-dependent. Most recently, the effect of tensor force on the formation of proton bubble structure in the spherical state of $^{46}$Ar has been studied with the Hartree-Fock-Bogoliubov (HFB) approach using either a Skyrme force~\cite{Wang11} or a semirealistic $NN$ interaction~\cite{Nakada13}. It has been found in Ref.~\cite{Wang11} that the proton bubble structure in $^{46}$Ar is possible if there is a strong inversion of 2s$_{1/2}$ and 1d$_{3/2}$ orbital induced by the tensor force. However, an opposite conclusion was drawn in Ref.~\cite{Nakada13} that the proton bubble structure is unlikely to be observed in any of the argon isotopes due to the strong anti-bubble effect of pairing correlations.  Therefore, the existence of bubble structure in $^{46}$Ar remains an open question. The new generation of electron-RIB collider SCRIT (Self-Confining Radioactive Isotope Target in Japan) under construction at RIKEN is able to measure density distribution of short-lived nuclei~\cite{Suda09} and is planned to settle down this question in the near future~\cite{Suda12}.

%

 Actually, there is another kind of correlations that might affect the density profile in $^{46}$Ar, i.e. dynamical quadrupole shape effects. These dynamical correlation effects are composed of two parts: (1) restoration of rotational symmetry for intrinsic quadrupole deformed states; (2) fluctuation in quadrupole shape degree of freedom. The former can shift the global minimum on the energy surface and therefore change the configuration of energy favored state. The latter leads to the spreading of the ground state wave function around the energy favored configuration. Recently, these effects have been examined on the bubble structure in low-lying states of $^{34}$Si within the framework of a particle-number and angular-momentum projected generator coordinate method (GCM+PNAMP) based on the mean-field approaches using either the non-relativistic Skyrme force SLy4~\cite{Yao12} or the relativistic energy density functional (EDF) PC-PK1~\cite{Yao13}. Both studies have demonstrated that the dynamical correlation effects can quench, but not smooth out completely the proton bubble structure in the ground state of $^{34}$Si.

 The aim of this work is to provide a beyond relativistic mean-field (RMF) study of the low-lying states and bubble structure in $^{46}$Ar. The reliability of the approach for low-lying spectroscopy is demonstrated by comparing with available data. The dynamical correlation effects on the proton bubble structure in $^{46}$Ar and other argon isotopes are examined.

 {\em The method}.- The wave function of nuclear low-lying state is given by the superposition of a set of both particle-number and angular-momentum projected axially deformed mean-field states
 \begin{equation}
 \vert \Psi^{JNZ}_\alpha\rangle
 =\sum_\beta f^{JNZ}_\alpha(\beta)\hat P^J_{M0} \hat P^N\hat P^Z\vert \Phi(\beta)\rangle,
 \end{equation}
 where $\hat P^J_{M0}$, $\hat P^N$, $\hat P^Z$ are the projection operators onto angular momentum, neutron and proton numbers, respectively. $\vert \Phi(\beta)\rangle$s are  axially deformed states from the RMF+BCS calculation with a constraint on the mass quadrupole moment $\langle Q_{20} \rangle = \sqrt{\dfrac{5}{16\pi}}\langle \Phi(\beta)\vert 2 z^2 - x^2 - y^2\vert  \Phi(\beta) \rangle$, where the deformation parameter $\beta$ is related to the quadrupole moment by $\beta= \dfrac{4\pi}{3AR^2} \langle Q_{20} \rangle$, $R=1.2A^{1/3}$ with mass number $A$.  Minimization of nuclear total energy with respect to the coefficient $f^{JNZ}_\alpha$ leads to the Hill-Wheeler-Griffin equation, the solution of which provides the energy spectrum and all the information needed for calculating the electric multipole transition strengths~\cite{Ring80}. More detailed introduction to the method has been given in Refs.~\cite{Yao13,Yao10}.

The density distribution of nucleons in $r$-space corresponding to the state $\vert \Psi^{JNZ}_\alpha\rangle$ is given by~\cite{Yao12,Yao13}
\begin{eqnarray}
\label{eq_GCM:70}
\rho^{J\alpha}(\br)
  & = &   \sum_{\beta\beta'}  f^{JZN}_{\alpha} (\beta')f^{JZN}_\alpha(\beta)
  \sum_{\lambda}  (-1)^{2\lambda}Y_{\lambda0}(\hat \br) \nonumber\\
 &&\times
 \langle  J0,\lambda 0\vert J 0\rangle  \sum_{K_2}(-1)^{K_2}\langle JK_2,\lambda -K_2\vert J0\rangle \nonumber \\
        &   & \times
  \int d\hat\br^\prime \rho^{JK_2}_{\beta'\beta}(\br^\prime)Y^\ast_{\lambda K_2}(\hat \br^\prime),
\end{eqnarray}
where $\rho^{JK_2}_{\beta'\beta}(\br)$ is defined as
\begin{eqnarray}
  \label{rho_JKNZ}
  \rho^{JK_2}_{\beta'\beta}(\br)
 &\equiv& \dfrac{2J+1}{2} \int^\pi_0 d\theta\sin(\theta) d^{J\ast}_{K_20}(\theta)\nonumber\\
 &&\times
 \langle  \Phi(\beta^\prime) \vert \sum_{i}\delta(\br-\br_i) e^{i\theta\hat J_y} \hat P^{N}\hat P^{Z}  \vert \Phi(\beta)\rangle.
\end{eqnarray}
 The index $i$ in the summation runs over all the occupied single-particle states for neutrons or protons. We note that the density by Eq.(\ref{eq_GCM:70}) contains the information of many deformed mean-field states generated by the collective coordinate $\beta$ and it corresponds to the density in the laboratory frame. For the ground state $0^+_1$, the density is simplified as
 \begin{eqnarray}
 \rho^{gs}(\br) =  \sum_{\beta\beta'}  f^{0ZN}_1 (\beta')f^{0ZN}_1(\beta) \int d\hat\br\rho^{00}_{\beta'\beta}(r, \hat\br),
 \end{eqnarray}
 where $\hat \br$ denotes the angular part of coordinate $\br$.

{\em Numerical details}.$-$ In the constrained mean-field calculations, parity, $x$-simplex symmetry, and time-reversal invariance are imposed. The Dirac equation is solved by expanding the Dirac spinor in terms of three-dimensional harmonic oscillator basis within 10 major shells. We adopt two popular parameterizations of relativistic point-coupling EDF PC-PK1~\cite{Zhao10} and PC-F1~\cite{Burvenich02}, in which, pairing correlations between nucleons are treated with the BCS approximation using a density-independent $\delta$ force implemented with a smooth cutoff factor.

\begin{figure}[]
\centering
\includegraphics[width=8cm]{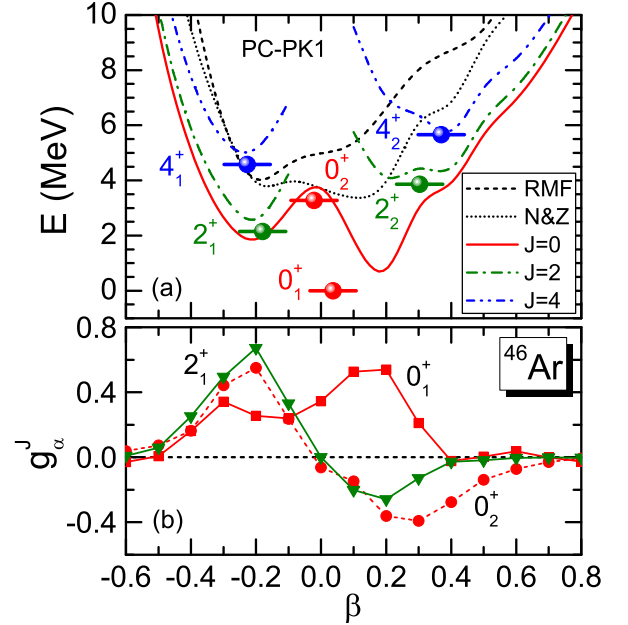}
\caption{(Color online) (a) Deformation energy curve from the constrained RMF+BCS calculation and those with additional projection onto particle number ($N\&Z$), and angular momentum ($J=0, 2, 4$), together with the final GCM states, which are placed at their average deformation. (b) Collective wave functions of $0^+_1, 0^+_2$ and $2^+_1$ states. All the results are calculated using the PC-PK1 force.}
\label{PECs}
\end{figure}

\begin{figure}[]
\centering
\includegraphics[width=8.5cm]{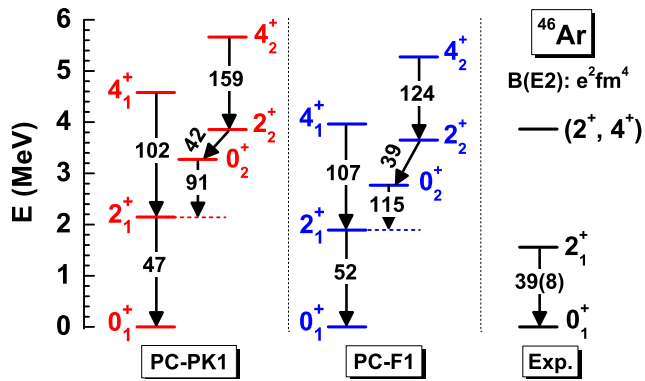}
\caption{(Color online) The spectra of $^{46}$Ar calculated with the GCM+PNAMP using the
PC-PK1 and PC-F1 forces, in comparison with available data~\cite{Scheit96,NNDC}.
The numbers on the arrows are the $B(E2)$ values (in units of $e^{2}$ fm$^{4}$). }
\label{spectrum}
\end{figure}

 {\em Spectroscopy of low-lying states}- Figure~\ref{PECs} displays the deformation energy curves with projection onto particle number $N, Z$ and additional angular momentum $J$. The low-lying states and the collective wave functions for the first three states from configuration mixing calculation are also plotted. The $g_\alpha^J(\beta)$s are related to the weight function $f^{JNZ}_\alpha$ by the following relation
\begin{equation}
\label{eq_GCM:30}
g_\alpha^{J}(\beta)
= \sum_{\beta'} [ \mathcal{N}^{J}(\beta,\beta') ]^{1/2} f^{JNZ}_\alpha(\beta'),
\end{equation}
 which provide the information of dominated configurations in the collective states. $\mathcal{N}^{J}(\beta,\beta')$ is the norm kernel defined by $\mathcal{N}^{J}(\beta,\beta')=\langle \Phi(\beta)\vert \hat P^J_{00} \hat P^N\hat P^Z\vert \Phi(\beta')\rangle$.

 We note that the mean-field energy curve exhibits a pronounced oblate deformed minimum with $\beta\simeq-0.2$, which is in agreement with our previous results from the triaxial relativistic Hartree-Bogoliubov calculation~\cite{Li11} using the DD-PC1~\cite{Niksic08} EDF for the particle-hole channel and a separable pairing force~\cite{Tian09} for the particle-particle channel. However, Fig.~\ref{PECs} shows that the PNP changes the energy surface significantly. More energy is gained in the weakly prolate deformed states than the weakly oblate states from the PNP, which shifts the global minimum to a weakly prolate deformed state. After projected onto angular momentum $J=0$, the global minimum is found at the prolate deformed state with $\beta\simeq+0.2$. Besides, the oblate deformed minimum with $\beta\simeq-0.2$ becomes an excited local minimum, which might be a saddle point if the triaxiality is taken into account. Our previous studies have shown that the triaxiality effect has only small influence on the excitation energy of $2^+_1$ state and $B(E2: 2^+_1\to 0^+_1)$ for nuclei of this mass region~\cite{Yao10}. Moreover, the GCM+PNAMP calculation with triaxiality evolves heavy numerical calculations. Therefore, triaxiality is not taken into account.

 It is seen in the panel (b) of Fig.~\ref{PECs} that the first two $0^+$ states are mixing of oblate and prolate deformed states at $\vert \beta\vert\simeq0.2$. Moreover, we obtain oblate deformed $2^+_1, 4^+_1$ states coexisting with prolate deformed $2^+_2, 4^+_2$ states in $^{46}$Ar. The results are similar to those by the GCM+AMP calculation (without PNP) using the D1S force~\cite{Guzman02}. The spectra of $^{46}$Ar calculated with the GCM+PNAMP using both the PC-PK1 and PC-F1 forces are compared with available data in Fig.~\ref{spectrum}. Both forces predict very similar low-energy structure for $^{46}$Ar, reproducing the data of $B(E2:2^+_1\to 0^+_1)$ measured via intermediate-energy Coulomb excitation~\cite{Scheit96}, but overestimating the excitation energy of $2^+_1$ state. The calculated low-lying spectroscopy reflects the underlying shell structure. Our results suggest a slightly weakening of the $N=28$ shell in $^{46}$Ar. The obtained neutron $N=28$ spherical energy gap is 4.09 (3.73) MeV by the PC-PK1 (PC-F1) force, in comparison with 4.80 MeV in $^{49}$Ca and 4.47 MeV in $^{47}$Ar, obtained by neutron stripping reactions~\cite{Gaudefroy06}.

\begin{figure}[]
\centering
\includegraphics[width=8.5cm]{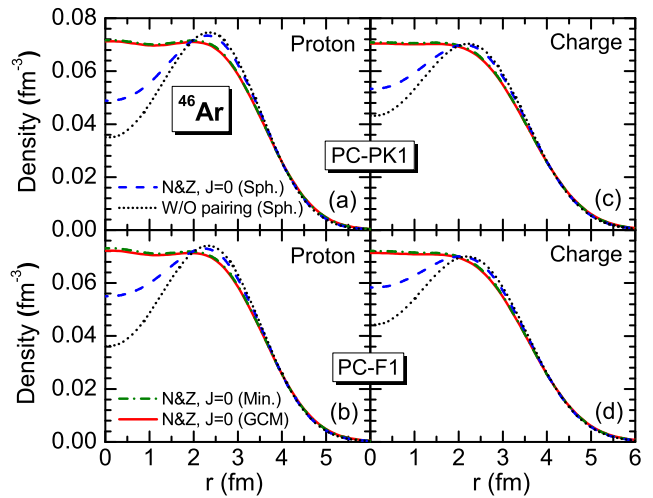}
\caption{(Color online) Proton and charge densities of $^{46}$Ar calculated using both PC-PK1 and PC-F1 forces. The densities from the RMF calculation without pairing (w/o pairing) is also given for comparison. More details are given in the text.}
\label{dens}
\end{figure}

 \begin{table}[b]
 \caption{The single-particle energies (in units of MeV) of  proton and neutron 2p$_{3/2}$ and 2p$_{1/2}$ orbitals and their splitting in the spherical state of $^{46}$Ar from the RMF calculation using the PC-PK1 force  with (w) and without (w/o)  the pairing correlations. }
 \begin{tabular}{c|ccc|ccc}
  \hline\hline
             &  \multicolumn{3}{c|}{proton} & \multicolumn{3}{c}{neutron} \\
 \hline
 Pairing      &  $\epsilon(2p_{3/2})$ &  $\epsilon(2p_{1/2})$ &  $\Delta\epsilon$ &  $\epsilon(2p_{3/2})$ &  $\epsilon(2p_{1/2})$ &  $\Delta\epsilon$  \\
 \hline
 w     &    $-2.622$            &      $-2.235$  & $-0.387$   &    $-3.704$          &     $-3.139$               & $-0.565$ \\
 w/o   &    $-2.221$            &      $-2.802$  & $+0.581$   &    $-3.240$          &     $-3.244$               & $+0.004$ \\
 \hline \hline
\end{tabular}
 \label{tab1}
\end{table}

\begin{figure}[htp]
\includegraphics[width=8cm]{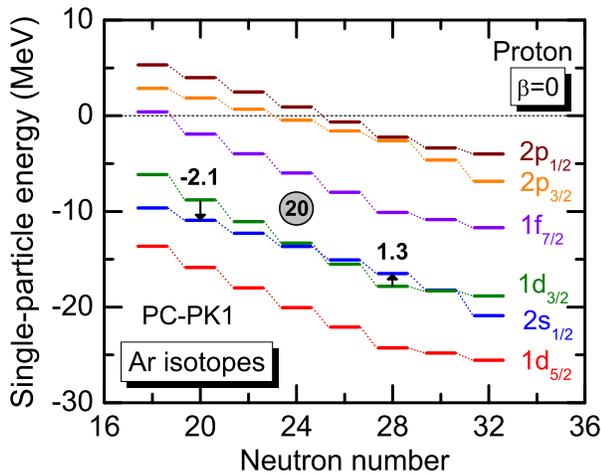}
 \caption{(Color online) The single-particle energy of protons in the spherical state of argon isotopes from the RMF+BCS calculation using the PC-PK1 force. }
\label{spe}
\end{figure}

 {\em Density profile and shell structure}- Figure~\ref{dens} displays the proton and charge densities of $^{46}$Ar from both mean-field and beyond mean-field calculations using both PC-PK1 and PC-F1 forces. A large depletion at $r=0$ (i.e., semi-bubble structure) is shown in both the proton and charge densities corresponding to the spherical state, in particular in the case of no pairing correlation. The inclusion of pairing correlation quenches but not eliminates the bubble structure.

 The persistence of bubble structure in the spherical state with pairing correlation can be understood from the underlying single-particle structure, as shown in Fig.~\ref{spe}. The inversion of 2s$_{1/2}$ and 1d$_{3/2}$ orbital is found in $^{44,46,48}$Ar. The single-particle energy difference between these two proton states $\Delta\epsilon_{13}=\epsilon (2s_{1/2})-\epsilon (1d_{3/2})$ is $-2.1$ MeV and $+1.3$ MeV in $^{38}$Ar and $^{46}$Ar, respectively, which is consistent with the corresponding data of $^{40}$Ca and $^{48}$Ca~\cite{Doll76,Ogilvie87}. In other words, the inversion of 2s$_{1/2}$ and 1d$_{3/2}$ orbital has been reproduced automatically in the spherical RMF calculation without the tensor force. It is different from the case found in the non-relativistic HFB calculations using several Skyrme EDFs (except the SkI5 force as demonstrated in Ref.~\cite{Khan08}), in which, one usually has to introduce the tensor force to reproduce the inversion~\cite{Wang11}. Since the $\Delta\epsilon_{13}$ reaches the largest value at $N=28$ in Ar isotopes, the largest central depletion among argon isotopes is shown in $^{46}$Ar.

 We note that the strength of spin-orbit interaction in the RMF approach is determined by the derivative of the potential $V(\br)-S(\br)$~\cite{Meng06}, where $V(\br)$ and $S(\br)$ are vector and scalar potentials respectively. A large central depletion
 in the potential $V(\br)-S(\br)$ is also found in the spherical state of $^{46}$Ar. As a result, the splitting of spin-orbit partners located mainly at the nuclear center, i.e. 2p$_{3/2}$-2p$_{1/2}$ is significantly reduced or even changes its sign in the case of no pairing, as shown in Tab.~\ref{tab1}. It is consistent with the conclusion in Ref.~\cite{Todd-Rutel2004} that the dramatic decrease in the spin-orbit splitting of $^{46}$Ar is not caused by the neutron density near the nuclear surface, but rather by the proton density in the nuclear interior.

\begin{figure}[]
\includegraphics[width=8cm]{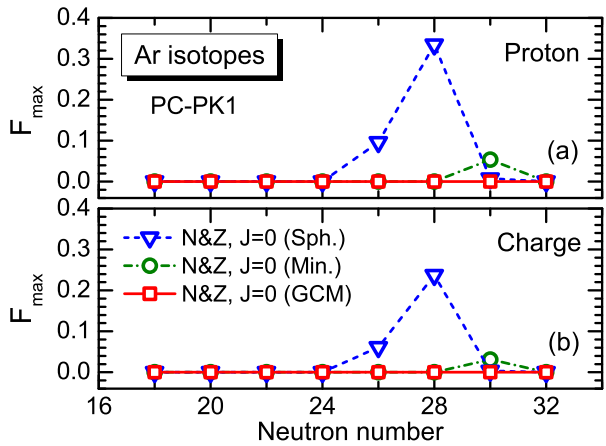}
\caption{(Color online) The depletion factor of proton (a) and charge (b) densities in argon isotopes by the PC-PK1 force. }
\label{factor}
\end{figure}

  The dynamical correlation effect on the bubble structure is demonstrated in Fig.~\ref{dens}, which also presents the densities corresponding to the state of the minimum on $J=0$ energy curve given by $\displaystyle\int d\hat\br\rho^{00}_{\beta_{\rm min.}\beta_{\rm min.}}(\br)$ and the ground state from full GCM calculation. The deformation parameter is $\beta_{\rm min.}=0.2$ and $\rho^{00}_{\beta,\beta}$ has been given in Eq.(\ref{rho_JKNZ}). The bubble structure disappears in the densities from the projection and additional GCM calculations, which include the dynamical deformation effects associated with AMP and shape mixing. We note that the dynamical correlation effect from the AMP plays a major role in quenching the bubble structure in most cases.

  Figure~\ref{factor} displays the depletion factor $F_{\text{max}}$ in the argon isotopes as a function of neutron number, where $F_{\text{max}} \equiv (\rho_{\textrm{max}}-\rho_{\textrm{cent}})/ \rho_{\textrm{max}}$ with $\rho_{\textrm{max}}$ being the largest value of the density in coordinate space and $\rho_{\textrm{cent}}$ the value at the center $r=0$. It is seen that the $F_{\text{max}}$ is zero for all the argon isotopes when the dynamical deformation effects are taken into account in the full GCM calculation. We have also checked the proton density in very neutron-rich $^{68}$Ar. Similar phenomenon has been found. Therefore, we conclude that the existence of proton bubble structure in argon isotopes is unlikely.

 {\em Summary}.- We have investigated the low-energy structure and anti-bubble effect of dynamical correlations associated with quadrupole shape in $^{46}$Ar by employing our newly established beyond RMF approach implemented with the GCM+PNAMP.  The low-spin energy spectrum, electric quadrupole transition strengths, and charge density in laboratory frame have been calculated. Our results are in excellent agreement with available data and suggest a slightly weakening of the $N=28$ shell in $^{46}$Ar. The inversion of 2s$_{1/2}$ and 1d$_{3/2}$ orbital has been found in the spherical states of $^{44,46,48}$Ar, which gives rise to a semi-bubble structure in the proton and charge densities. However, this bubble structure eventually disappears after taking into account the effect of dynamical correlations. Our results indicate that the observation of proton bubble structure in argon isotopes is unlikely. These findings are hoped to be examined on the SCRIT facility in the near future~\cite{Suda12}.

  We note that in the present calculations, the tensor contribution is completely absent due to the missing of Fock terms. The inclusion of tensor contribution within the relativistic Hartree-Fock approach (RHF)~\cite{Long07} may enlarge the energy difference between 2s$_{1/2}$ and 1d$_{3/2}$ orbital~\cite{Torres10} and acts in the opposite direction with respect to the dynamical correlations on the proton bubble structure. In other words, the tensor force may revive the bubble structure. Therefore, the study of tensor contribution to the proton bubble structure in neutron-rich Ar isotopes within the RHF approach will be very interesting.

 {\em Acknowledgments}.- We thank T. Suda for drawing us attention to the density distribution in $^{46}$Ar and thank K. Hagino and H. Mei for valuable discussions. This work was supported in part by the Tohoku University Focused Research Project ``Understanding the origins for matters in universe", the NSFC under Grant Nos. 11305134, 11105110, 11105111, and 10947013, the Natural Science Foundation of Chongqing cstc2011jjA0376, and the Fundamental Research Funds for the Central Universities  (XDJK2010B007, XDJK2011B002 and XDJK2013C028).


\end{CJK}
\end{document}